\documentclass{emulateapj}
\usepackage{natbib}  
\usepackage{graphicx}
\usepackage{amsmath}
\usepackage{hyperref}
\usepackage{color}
\usepackage{xspace}
\usepackage[usenames,dvipsnames]{xcolor}

\newcommand{\msun}{\ensuremath{\mathrm{M}_{\odot}}\xspace}
\newcommand{\mdot}{\ensuremath{\dot{M}}\xspace}
\newcommand{\lsun}{L\ensuremath{_{\odot}}\xspace}
\newcommand{\kms}{\ensuremath{\mathrm{km~s}^{-1}}\xspace}

\newcommand{\rcluster}{2.5\xspace}
\newcommand{\nsample}{456\xspace}
\newcommand{\ncandidates}{18\xspace}

\newcommand{\nMPC}{3\xspace} 
\newcommand{\obsfrac}{30}
\newcommand{\nMPCtot}{10\xspace} 
\newcommand{\nMPCtoterr}{6\xspace} 

\newcommand{\mmin}{\ensuremath{10^4~\msun}\xspace}
\newcommand{\permyr}{\ensuremath{\mathrm{Myr}^{-1}}\xspace}

\newcommand{\percc}{\ensuremath{\mathrm{cm}^{-3}}\xspace}
\newcommand{\middistcut}{13.0\xspace}
\newcommand{\mindist}{8.7\xspace}
\newcommand{\um}{\ensuremath{\mu}m\xspace}

\newcommand{\CFR}{5\xspace} 
\def\ee#1{\ensuremath{\times10^{#1}}\xspace}
\newcommand{\tsuplim}{0.5\xspace} 

\slugcomment{To be submitted to ApJL:  DRAFT -   \today }
\shorttitle{No Starless Massive Proto-Clusters}
\shortauthors{A. Ginsburg  et al.}

\begin{document}

\title{There are no starless massive proto-clusters in the first quadrant of the Galaxy}

\author{
  A. Ginsburg\altaffilmark{1},
  E. Bressert\altaffilmark{2,3},
  J. Bally\altaffilmark{1},
  C. Battersby\altaffilmark{1}}

\altaffiltext{1}{Center for Astrophysics and Space Astronomy, 
     University of Colorado, Boulder, CO 80309}
\altaffiltext{2}{European Southern Observatory, Karl Schwarzschild str. 2, 
     85748 Garching bei M\" unchen, Germany} 
\altaffiltext{3}{School of Physics, University of Exeter, 
     Stocker Road, Exeter EX4 4QL, UK}

\begin{abstract} 
    
We search the $\lambda=1.1$ mm Bolocam Galactic Plane Survey for clumps
containing sufficient mass to form $\sim10^4~\msun$ star clusters.
\ncandidates\ candidate massive proto-clusters  are identified in the first Galactic quadrant outside
of the central kiloparsec.  This
sample is complete to clumps with mass M$_{\rm clump}>\mmin$ and radius
$r\lesssim2.5$ pc.  The overall Galactic massive cluster formation rate is
$CFR({\rm M}_{\rm cluster}>10^4) \lesssim \CFR\  \permyr$, which is in
agreement with the rates inferred from Galactic open clusters and M31 massive
clusters.  We find that all massive proto-clusters in the first quadrant are
actively forming massive stars and place an upper limit of
$\tau_{starless}<\tsuplim$~Myr on the lifetime of the starless phase of massive
cluster formation.  If massive clusters go through a starless phase with all 
of their mass in a single clump, the lifetime of this phase is very short.

\end{abstract}


\section{Introduction}

The Milky Way contains about 150 Globular clusters (GCs) with masses of $10^4$
to over $10^6$ \msun\ and tens of thousands of open clusters containing from
100 to over $10^4$ stars.  However, young massive clusters containing
$\gtrsim10^4~\msun$ of stars are rare, with only a handful known
\citep{PortegiesZwart2010}. While no GCs have formed in the Milky Way within
the last 5 Gyr, open clusters that survive many crossing times continue to
form.   A few of these clusters have stellar masses greater than $10^4$
M$_{\odot}$ and therefore qualify as young massive clusters
\citep[YMCs;][]{PortegiesZwart2010}.   YMCs must either form from clumps having
masses greater than and sizes comparable to the final cluster  or be formed
from a larger, more diffuse reservoir, in which case massive protocluster
clumps may be rare or nonexistent  \citep{Kennicutt2012}.

Massive proto-clusters (MPCs) are massive clusters (M$_{\rm cluster}>10^4$ \msun)
in the process of forming from a dense gas cloud.  In \citet{Bressert2012}, we
examine the theoretical properties of MPCs: MPCs are assumed to form from
massive, cold starless clumps analagous to pre-stellar cores
\citep{Williams2000}.  In this paper, we refer to two classes of objects:
starless MPCs, which have very low luminosity and do not contain OB stars, and
MPCs, which are gas-rich but have already formed OB stars.  The only
currently known starless MPC is G0.253+0.016, which lies within the dense
central molecular zone and is subject to greater environmental stresses than
similar objects in the Galactic plane \citep{Longmore2012}.

Because massive clusters contain many massive stars, at some point during their
evolution ionization pressure will prevail over protostellar outflows as the
dominant feedback mechanism.  Other sources of feedback are less than
ionization pressure up until the first supernova explosion
\citep{Bressert2012}.  These proto-clusters must have masses
M$_{\rm clump}>{\rm M}_{*}/SFE$ \footnote{We define a star formation efficiency
$SFE={\rm M}_{\rm *,final} / {\rm M}_{\rm gas,initial}$.}, or about $3\ee{4}$ \msun\ for an assumed
SFE=30\% (an upper limit on the star formation efficiency),
confined in a radius $r\lesssim2.5$ pc, in order to remain bound against
ionization feedback.  These properties motivate our search for proto-clusters
in the Bolocam Galactic Plane Survey \citep[BGPS;][
\url{http://irsa.ipac.caltech.edu/data/BOLOCAM_GPS/}]{Aguirre2011}.

The distinction between relatively short-lived `open clusters' and long-lived
($t\gtrsim1$ Gyr) bound clusters occurs at about $10^4$ \msun
\citep{PortegiesZwart2010}.  Clusters with ${\rm M}_{\rm cluster} < 1\ee{4} \msun$~will
be destroyed by interactions with giant molecular clouds over the course of a
few hundred million years after they have dispersed their gas
\citep{Kruijssen2011}, while clusters with ${\rm M}_{\rm cluster}\gtrsim10^4 \msun$ may
survive $\gtrsim 1$ Gyr.  Closer to the Galactic center, within approximately a
kiloparsec, all clusters will be destroyed on shorter timescales by strong
tidal forces or interactions with molecular clouds.

In the Galaxy, there are few known massive clusters.
\citet{PortegiesZwart2010} catalogs a few of them, of which NGC 3603, Trumpler
14, and Westerlund 1 and 2 are the likely descendants of the objects we
investigate.  These clusters have $r_{eff} \lesssim 1$ pc, $M\sim10^4$ \msun,
and ages $t\lesssim4$ Myr.  We present a census of their ancestral analogs.

\section{Observations and Analysis}
\label{sec:observations}

\subsection{The Bolocam Galactic Plane Survey}
\begin{figure*}
    \includegraphics[width=7in]{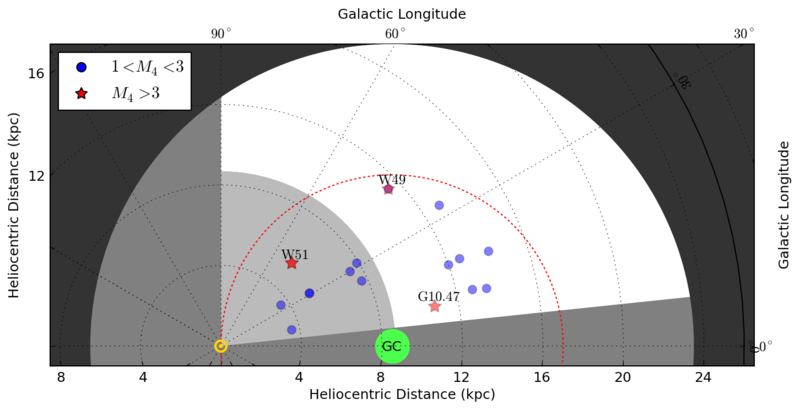}
\caption{
\label{fig:galplot}
Plot of the massive proto-cluster (MPC) candidates
overlaid on the Galactic plane.  
The green circle represents the galactic center, and the yellow $\odot$ is the
Sun.
A 15 kpc radius disc centered on the Galactic Center indicates the approximate
extent of Galactic star formation.  The white region indicates the coverage of
the
Bolocam Galactic Plane survey and our source selection limits based on distance
and longitude.  The inner cutoff (light grey) is the nearby incompleteness
limit set by the Bolocam spatial filtering;  the catalog includes sources but
is incomplete in this region.  The red dashed circle traces the solar circle.
Blue filled circles represent initial candidates that passed the mass-cutoff
criterion $M(20K)>\mmin$; red stars are those with $M(20K) > 3\ee{4} \msun$.
In the legend, $M_4$ means mass in units of $10^4 \msun$.  
}
\end{figure*}

The BGPS is a 1.1 mm survey of the first quadrant of the Galactic plane in the
range $-0.5 < b < 0.5$ with resolution $\sim33\arcsec$ sensitive to a maximum
spatial scale of $\sim120\arcsec$ \citep{Aguirre2011,Ginsburg2012}.  The BGPS `Bolocat' v1.0 catalog
includes sources identified by a watershed decomposition algorithm and flux
measurements within apertures of radius 20\arcsec, 40\arcsec, and 60\arcsec\
\citep{Rosolowsky2010}.

We searched the BGPS for candidate MPCs in the 1st quadrant ($6 < \ell < 90$;
5991 sources).  The inner 6 degrees of the Galaxy are excluded because physical
conditions are significantly different from those in the rest
of the galaxy  \citep{YusefZadeh2009} and the BGPS is confusion-limited in 
that region.

\subsection{Source Selection \& Completeness}
\label{sec:selection}
We identify a flux-limited sample by setting our search criteria to
include all sources with ${\rm M}_{\rm clump}>10^4$ \msun\ in a 20\arcsec\ radius out to 26 
kpc, or a physical radius of 2.5 pc at that distance.  The radius cutoff is
motivated by completeness and physical considerations: the cutoff of 26 kpc includes
the entire star forming disk in our targeted longitudes, and $r=2.5$ pc corresponds
to the radius at which a $3\ee{4}$ \msun\ mass has an escape speed $v_{esc}=10$ \kms, i.e.
ionized gas will be bound. 
The maximum radius and minimum mass imply a minimum mean density
$\bar{n}=6\times10^3~\percc$, which implies a maximum free-fall time $t_{ff}<0.65$~Myr.

Using the Bolocat v1.0 catalog, we first set a flux limit on the sample by assuming
the maximum distance of $d=26$ kpc and imposing a mass cutoff of ${\rm M}_{\rm clump}\geq10^4$ \msun\ 
inside a 20\arcsec\ (2.5 pc) radius aperture.  Following equation 19 in
\citet{Aguirre2011}:
\begin{equation} 
    {\rm M}_{\rm gas}\approx 14.3 \left( e^{13.0/T_d}-1 \right)
        \left({S_\nu\over 1\; {\rm Jy}} \right)
        \left(\frac{D}{{\rm 1~kpc}}\right)^{2} \msun 
\end{equation}   
and assuming $T_{dust}=20$K, the implied flux cutoff is 1.13 Jy \footnote{As
per \citet{Rosolowsky2010}, \citet{Aguirre2011}, and \citet{Ginsburg2012}, a
factor of 1.5 calibration correction and 1.46 aperture correction are required
for the 20\arcsec\ radius aperture fluxes reported in the catalog.  These
factors have been applied to the data. }, above which \nsample\ `flux-cutoff'
candidates were selected in the Bolocat v1.0 catalog.  Cutoffs of 4.3 Jy for
the 40\arcsec\ and 10.2 Jy for the 60\arcsec\ Bolocat v1.0 apertures were used
to select more nearby candidates inside the same physical radius, but no
sources were selected based on these larger apertures.

The BGPS is insensitive to scales larger than 120\arcsec\
\citep[][]{Ginsburg2012}\footnote{\citet{Ginsburg2012} presents v2.0 of the
BGPS}.  As a result, the survey is incomplete below a distance $$D_{min} =
\mindist \left(\frac{r_{\rm cluster}}{2.5 \textrm{pc}}\right) \textrm{kpc} $$ from
the Sun.  Within this radius, alternate methods must be sought to determine the
total mass within $r_{\rm cluster} < \rcluster$ pc.  Although the sample is
incomplete for $D < \mindist$ kpc, sources that have sufficient mass despite
the 120\arcsec\ spatial filtering are included.

Distances to BGPS-selected candidates were determined primarily via literature
search.  Where distances were unavailable, we used velocity measurements from
\citet{Schlingman2011} and assumed the far distance for source selection.  We
then resolved the kinematic distance ambiguity towards these sources by
searching for associated near-infrared stellar extinction features from the
UKIDSS GPS \citep{Lucas2008}.
Most literature distances were determined using a
rotation curve model and some method of kinematic distance ambiguity
resolution. Because the literature used different rotation curve models, there
is a $\sim10\%$ systematic error in distance resulting in a $\sim20\%$
systematic error in mass. We used the larger
40\arcsec\ radius apertures to determine the flux for sources at
$D<\middistcut$ kpc and 60\arcsec\ radius apertures for sources at $D<\mindist$
kpc (corresponding to $r<\rcluster$ pc).

The masses were computed assuming a temperature $T_{dust}=20$K, opacity
$\kappa_{271.1 GHz} = 0.0114~\mathrm{cm}^2 \mathrm{g}^{-1}$, and gas-to-dust
ratio of 100  \citep{Aguirre2011} \footnote{$T_{dust}=20$K is more appropriate
for a typical pre-star-forming clump than an evolved HII-region hosting one
\citep[e.g.]{Dunham2010}. However, because we are interested in cold
progenitors as well as actively forming clusters, the selection is based on
$T_{dust}=20$K, which is more inclusive. }.  The mass estimate drops by a
factor of $2.38$ if the temperature assumed is doubled to $T_{dust}=40$K.  

\citet{Ginsburg2011} notes that significant free-free contamination, as high as
80\%, is possible for some 1.1 mm sources, so the selected candidates may prove
to be more moderate-mass and evolved proto-clusters.  We used the NRAO VLA
Archive Survey \citep[NVAS;][]{Crossley2008} to estimate the free-free
contamination for the sample.  For most sources, the free-free contamination
inferred from the VLA observations is small ($<20\%$), but for a subset the
contamination was $\sim20-35\%$ assuming that the free-free emission is
optically thin.  Corrected masses using the measured free-free contamination
and higher dust temperatures are listed in Table \ref{tab:candidates}; these
are reasonable lower limits on the total mass of these regions.  All of the
contamination estimates are technically lower limits both because of the
assumption that the free-free emission is optically thin and because the VLA
filters out large-scale flux.  However, in most cases, the emission is likely
to be dominated by optically thin emission \citep[evolved HII regions tend to
be optically thin and bright, while compact HII regions are optically thick but
relatively faint;][]{Keto2002} and for most sources VLA C or D-array
observations were used, and at 3.6 and 6 cm the largest angular scale recovered
is 180-300 \arcsec, greater than the largest angular scale in the BGPS.  

Applying a cutoff of M$_{\rm clump} > 10^4$ \msun\ left \ncandidates\
protocluster candidates out of the original \nsample.  The more stringent cut
M$_{\rm clump} > 10^4 / SFE \approx 3\ee{4}$ \msun\ leaves only \nMPC\ MPCs . 

The final candidate list contains only sources with $M(20K)>10^4 \msun$ (the
completeness limit; see Table \ref{tab:candidates}).  The table lists
their physical properties, their literature distance, their mass (assuming $T_{dust}=20
\textrm{~and~} 40 K$ and a free-free subtracted lower-limit) ,
and their inferred escape speed ($v_{esc} = \sqrt{2 G M(20K) / r}$) assuming a
radius equal to the aperture size at that distance.  The table also includes
measurements of the IRAS luminosity in the 60 and 100 \um\ bands within the
source aperture.

\subsection{Source Separation}
These \ncandidates\ candidates include some overlapping sources.
There are two clumps in W51 separated by about 1.5 pc and 4.5 \kms\ along the
line of sight that are each independently massive enough to be classified as
MPCs, but are only discussed as a single entity because they are likely
to merge if their three-dimensional separation is similar to their projected
distance.  The candidates in W49 are more widely separated, about 4.4 pc and 7
\kms\ along the line of sight, but could still merge.

Additionally, 9 of the \ncandidates\ are within 8.7 kpc, so the mass
estimates are lower limits.  These are promising candidates for follow-up, but
cannot be considered complete for population studies.  If our radius restriction
is dropped to 1.5 pc, the minimum complete distance drops to 5.2 kpc and the
three lowest-mass sources in Table \ref{tab:candidates} no longer qualify, but
otherwise the source list remains unchanged.  Our analysis is therefore robust
to the selection criteria used.

\section{Results}

\subsection{Cluster formation rate}
\label{sec:cfr}

The massive clumps in Table \ref{tab:candidates} can be used to constrain the
Galactic formation rate of massive clusters (MCs) above \mmin\ if we assume
that the number of observed proto-clusters is a representative sample. The region
surveyed covers a fraction of the surface area of the Galaxy
$f_{observed}=A_{survey} / A_{Galaxy} \approx \obsfrac\%$ assuming the star
forming disk has a radius of 15 kpc\footnote{The observed fraction of the
galaxy changes to 21\% if we only include the area within the solar
circle as discussed in \S \ref{sec:discussion}.}.
The cumulative
cluster formation rate above a cluster mass ${\rm M}_{cl}$ is given by $$CFR(>{\rm M}_{cl})
= \frac{N_{MPC}}{\tau_{SF} f_{observed}}$$ where $ \tau_{SF} \approx 2$\ Myr is
the assumed cluster formation timescale \footnote{$\tau_{SF}$, the time from the start of star formation
until gas expulsion, is a poorly understood
quantity, but is reasonably constrained to be $\gtrsim1$~Myr from the age
spread in the Orion Nebula cluster \citep{Hillenbrand1997} and $\lesssim10$~Myr
because the most massive stars will go supernova by that time.}.
With the measured
$N_{MPC}({\rm M}_{\rm cluster}>10^4\msun) = \nMPC $\ proto-clusters, we infer a Galactic formation rate 
$$CFR \lesssim \CFR \left(\frac{\tau_{SF}}{2
~\textrm{Myr}}\right)^{-1} \textrm{~Myr}^{-1}$$  This cluster formation rate is
statistically weak, with Poisson error of about 3.5 
\permyr\ and can be improved with more complete surveys \citep[e.g., Hi-Gal,][]{Molinari2010}.  This
formation rate is an upper limit because all of the estimated
masses are upper limits as discussed in Section \ref{sec:selection}.

\subsection{Comparison to Clusters in Andromeda}

We use cluster observations in M31 from \citet{Vansevicius2009} to infer the
massive cluster formation rate in M31.  They observe 2 clusters with
${\rm M}_{\rm cluster}>10^4\msun$ and ages $<10$ Myr in 15\% of the M31 star-forming
disk.  The implied cluster formation rate in Andromeda is $\dot{N_{cl}} =
N_{cl}/0.15 / (10 ~\mathrm{Myr}) \approx 1.3$ \permyr.  Given M31's total star
formation rate $\sim 5\times$ lower than the Galactic rate \citep[Andromeda
$\mdot=0.4$, Milky Way $\mdot=2$ \msun \permyr;][]{Barmby2006,Chomiuk2011}, the
predicted Galactic cluster formation rate is $\dot{N_{cl}}(MW) = 5~
\dot{N_{cl}}(M31) = 6.5$ \permyr \citep[assuming the CFR scales linearly with
the SFR; ][]{Bastian2008}.  
The scaled-up Andromeda cluster formation rate matches the observed Galactic
cluster formation rate.  The samples are small, but as a sanity check, the
agreement is comforting.

\subsection{Star Formation Activity}

In the sample of potential proto-clusters, all have formed massive stars based
on a literature search and IRAS measurements.  A few of the low mass sources,
G012.209-00.104, G012.627-00.016, G019.474+00.171, and G031.414+00.307 have
relatively low IRAS luminosities ($L_{IRAS} = L_{100}+L_{60} < 10^5 \lsun$) and
little free-free emission.  However, \emph{all} are detected in the radio as
H~II regions (some ultracompact) and have luminosities indicating early-B type
powering stars.

Non-detection of `starless' proto-cluster clumps implies an upper limit on the
starless lifetime. For an assumed $\tau_{sf} \sim 2$~Myr, the $1\sigma$ upper
limit on the starless proto-MC clump is $\tau_{starless} <
(\sqrt{N_{cl}}/N_{cl}) \tau_{sf} = \tsuplim~\mathrm{Myr}$ assuming Poisson
statistics and using all 18 sources.  This limit is consistent with massive
star formation on the clump free-fall timescale ($\tau_{ff}\leq0.65$ Myr).  It
implies that massive stars form rapidly when these large masses are condensed
into cluster-scale regions and hints that massive stars are among the first to
form in massive clusters.

\section{Discussion}
\label{sec:discussion}

Assuming a lower limit 30\% SFE and T$_{dust} = 20 {\rm K}$, \nMPC\ candidates
in Table \ref{tab:candidates}  will become massive clusters like NGC 3603:
G010.472+00.026, W51, and W49 (G043.169+00.01).  Even if  T$_{dust} = 40 {\rm
K}$, W49 is still likely to form a $\sim10^4$ \msun\ MC, although G10.47 would
be too small.  W51, which is within the spatial-filtering incompleteness zone,
passes the cutoff and is likely to form a pair of massive clusters.  However,
if the dust in W51 is warm and the free-free contamination is considered, the
total mass in each of the W51 clumps is below the 3\ee{4} \msun\ cutoff.

The BGPS covers about \obsfrac\% of the Galactic star-forming disk in the range
1 kpc $< R_{gal}<15$ kpc.  We can extrapolate our \nMPC\ detections to predict
that there are $\leq$\nMPCtot\ ($\pm \nMPCtoterr$) proto-clusters in the Galaxy
outside of the Galactic center.
The agreement between the SFR-based prediction
from M31 and our observations implies that we have selected genuine massive
proto-clusters (MPCs).  

These most massive sources have escape speeds greater than the sound speed in
ionized gas, indicating that they can continue to accrete gas even after the
formation of massive stars.  Assuming they are embedded in larger-scale gas
reservoirs, we are measuring lower bounds on the `final' clump plus cluster
mass.

All of the young massive proto-clusters candidates observed are within the
solar circle despite our survey covering more area outside of
the solar circle.  
The outer radius limit for massive cluster formation is consistent with the
observed metallicity shift noted at the same radius by \citet{Lepine2011}.
They identify the solar circle as the corotation radius of pattern speed and
orbits within the Galaxy (within this radius, stars orbit faster than the
spiral pattern).  The fact that this radius also represents a cutoff between
the inner, massive-cluster-forming disk and the outer, massive-cluster-free
disk hints that gas crossing spiral arms may be the triggering mechanism for
massive cluster formation.  However, given the small numbers, the detected
clusters are consistent with a gaussian + exponential disk distribution
following that described by \citet{Wolfire2003}.

Future work should include a census for MPCs within $D\lesssim5$ kpc using the
Herschel Hi-Gal survey \citep{Molinari2010} and in the Southern plane with
ATLASGAL \citep{Schuller2009}.  Some surveys have already identified
proto-clusters in these regions \citep[e.g.][]{Faundez2004,Battersby2011}, but
they are not complete.  A complete survey of distances will be essential for
continuum surveys to be used.

There are two modes of massive cluster formation consistent with our
observations that can be observationally distinguished.  Either a compact
starless massive proto-cluster phase does occur and is short, or the mass to be
included in the cluster is accumulated from larger volumes over longer
timescales.  Extending our proto-cluster survey to the Southern sky, e.g. using
the ATLASGAL and Hi-Gal surveys, will either discover starless MPCs or
strengthen the arguments that there is no starless MPC phase.  If instead
massive clusters form by large scale ($r>2.5$ pc) accretion, substantial
reservoirs of gas should surround these most massive regions and be flowing
into them.  Signatures of this accretion process should be visible: MPCs should
contain molecular filamentary structures feeding into their centers
\citep[e.g.][]{Correnti2012,Hennemann2012,Liu2012}.  Alternatively, lower mass
clumps may merge to form massive clusters \citep{Fujii2012}, in which case
clusters of clumps - which should be detectable in extant galactic plane
surveys - are the likely precursors to massive clusters.  Finally, massive
clusters may form from the global collapse of structures on scales larger than
we have probed, which could also produce clusters of clumps.

\section{Conclusions}
\label{sec:conclusions}

Using the BGPS, we have performed the first flux-limited census of massive
proto-cluster candidates.  We found \ncandidates\ candidates that will be part
of the next generation of open clusters and \nMPC\ that could form massive
clusters similar to NGC 3603 (${\rm M}_{\rm cluster} > 10^4$ \msun).   We have
measured a Galactic massive cluster formation rate $CFR({\rm M}_{\rm
cluster}>10^4) \lesssim \CFR\  \permyr$\ assuming that clusters are equally
likely to form everywhere within the range 1 kpc $ < R_{gal} < $ 15   kpc. 
The observed MPC counts are
consistent with observed cluster counts in Andromeda scaled up by $SFR_{M31} /
SFR_{MW}$ assuming a formation timescale of 2 Myr.  

Despite this survey being the first sensitive to pre-star-forming MPC clumps, none
were detected.  This lack of detected pre-star-forming MPCs suggests a
timescale upper limit of about $\tau_{starless}<\tsuplim$ Myr for the pre-massive-star phase of
massive cluster formation, and hints that massive clusters may never form
highly condensed clumps ($\bar{n}\gtrsim10^4~\percc$) prior to forming massive
stars.
It leaves open the possibility that massive clusters form from large-scale
($\gtrsim 10$ pc) accretion onto smaller clumps over a prolonged ($\tau > 2$
Myr) star formation timescale.

Observations are needed to distinguish competing models for MC formation:
Birth from isolated massive proto-cluster clumps, either compact and rapid
or diffuse and slow, or from smaller clumps that
never have a mass as large as the final cluster
mass.  
This sample of the \ncandidates\ most massive proto-cluster clumps in the first
quadrant (where they can be observed by both the VLA and ALMA) presents an ideal
starting point for these observations.

\section{Acknowledgements}
We thank the referee for thorough and very helpful comments that strengthened
this Letter.  This work was supported by NSF grant AST 1009847.

\begin{table*}
\scriptsize
\begin{center}
\caption{\label{tab:candidates}
Massive Protocluster Candidates detected in the Bolocam Galactic Plane Survey with $M>10^4 \msun$ }
\begin{tabular}{ccccccccccc}
\hline
Name & Common & Distance & M(20K) & M(40K) & $^a$M(min) & Radius & $\bar{n}(H_2)$ & $v_{esc}$ & $^bf_{ff}$ & L(IRAS) \\
 & Name & kpc & $1000 M_{\odot}$ & $1000 M_{\odot}$ & $1000 M_{\odot}$ & pc & $10^4$cm$^{-3}$ & km~s$^{-1}$ &  & $10^5 L_{\odot}$ \\
\hline\hline
G010.472+00.026 & G10.47 & 10.8$^{7}$ & 38 & 16 & 16 & 2.1 & 1.4 & 12.7 & 0.01 & 5.0 \\
G012.209-00.104 & - & 13.5$^{7}$ & 14 & 6 & 5 & 1.3 & 2.3 & 9.9 & 0.05 & 0.61 \\
G012.627-00.016 & - & 12.8$^{9}$ & 10 & 4 & 3 & 2.5 & 0.2 & 5.9 & 0.05 & 0.59 \\
G012.809-00.200 & W33 & 3.6$^{7}$ & 12 & 5 & 3 & 1.0 & 3.8 & 10.2 & 0.32 & 3.0 \\
G019.474+00.171 & - & 14.1$^{12}$ & 11 & 4 & 4 & 1.4 & 1.6 & 8.6 & 0.02 & 0.26 \\
G019.609-00.233 & G19.6 & 12.0$^{7}$ & 26 & 11 & 7 & 2.3 & 0.7 & 10.0 & 0.31 & 6.4 \\
G020.082-00.135 & IR18253 & 12.6$^{10}$ & 13 & 5 & 4 & 2.4 & 0.3 & 6.8 & 0.14 & 2.8 \\
G024.791+00.083 & G24.78 & 7.7$^{11}$ & 14 & 6 & 5 & 2.2 & 0.4 & 7.4 & 0.11 & 1.5 \\
G029.955-00.018 & - & 7.4$^{3}$ & 10 & 4 & 2 & 2.2 & 0.3 & 6.4 & 0.34 & 5.3 \\
G030.704-00.067 & W43b & 5.1$^{6}$ & 11 & 4 & 4 & 1.5 & 1.1 & 8.0 & 0.11 & 1.0 \\
G030.820-00.055 & W43a & 5.1$^{10}$ & 11 & 4 & 4 & 1.5 & 1.2 & 8.1 & 0.13 & 1.9 \\
G031.414+00.307 & G31.41 & 7.9$^{2}$ & 18 & 7 & 7 & 2.3 & 0.5 & 8.3 & 0.05 & 0.8 \\
G032.798+00.193 & G32.80 & 12.9$^{1}$ & 22 & 9 & 7 & 2.5 & 0.5 & 8.9 & 0.27 & 6.9 \\
G034.258+00.154 & G34 & 3.6$^{4}$ & 13 & 5 & 4 & 1.0 & 4 & 10.5 & 0.12 & 2.7 \\
G043.164-00.031 & W49 & 11.4$^{5}$ & 24 & 10 & 6 & 2.2 & 0.7 & 9.7 & 0.38 & 9.9 \\
G043.169+00.009 & W49 & 11.4$^{5}$ & 120 & 52 & 39 & 2.2 & 4 & 22.2 & 0.25 & 16.0 \\
G049.489-00.370 & W51IRS2 & 5.4$^{8}$ & 48 & 20 & 14 & 1.6 & 4.3 & 16.2 & 0.27 & 4.5 \\
G049.489-00.386 & W51MAIN & 5.4$^{8}$ & 52 & 22 & 15 & 1.6 & 4.7 & 17.0 & 0.29 & 4.7 \\
\hline
\end{tabular}
\end{center}{\scriptsize 1: \citet{Araya2002}, 2: \citet{Churchwell1990}, 3: \citet{Fish2003}, 4: \citet{Ginsburg2011}, 5: \citet{Gwinn1992}, 7: \citet{Pandian2008}, 8: \citet{Sato2010}, 9: \citet{Sewilo2004}, 10: \citet{Urquhart2012}, 11: \citet{Vig2008}, 12: \citet{Xu2003}.  6: The distances to G030.704 was determined using the
near kinematic distance from the velocity of the HHT-observed HCO+ line \citep{Schlingman2011}.
$^a$: The minimum likely mass, $M_{min} = (1-f_{ff}) M(40K)$.
$^b$: The fraction of flux from free-free emission (as opposed to dust emission) at $\lambda=1.1$ mm
}
\end{table*}

\end{document}